\documentclass [aps, prl, amssymb, twocolumn, showpacs,superscriptaddress]{revtex4-1}
\usepackage{graphicx,epsfig, amsfonts, amssymb, amsmath}
\usepackage{bm}
\usepackage{times}
\usepackage[colorlinks,citecolor=blue,linkcolor=red]{hyperref}

\begin{document} 
 
\title{Braid Group and Topological Phase Transitions in Nonequilibrium Stochastic Dynamics}

\author {Jie Ren}\email{renjie@lanl.gov}
\affiliation{Theoretical Division, Los Alamos National Laboratory, Los Alamos, NM 87545} 
\author{N.~A. Sinitsyn}
\affiliation{Theoretical Division, Los Alamos National Laboratory, Los Alamos, NM 87545} 

\date{\today}

\begin{abstract}
We show that distinct topological phases of the band structure of a non-Hermitian Hamiltonian can be classified with elements of the braid group. As the proof of principle, we consider the non-Hermitian evolution of the statistics of nonequilibrium stochastic currents. We show that topologically nontrivial phases have detectable properties, including the emergence of decaying oscillations of parity and state probabilities, and discontinuities in the steady state statistics of currents.
\end{abstract}

\pacs{05.60.-k, 05.70.Ln, 64.70.-p, 82.37.-j}

\maketitle

Numerous physical systems are described by Hermitian Hamiltonians, $\hat{H}({\bm \chi})$, which periodically depend on a vector of parameters ${\bm \chi}$ with a period $2\pi$. As functions of ${\bm \chi}$, the eigenvalues $\lambda_i({\bm \chi})$ of $\hat{H}({\bm \chi})$ represent the {\it band structure} of the system. Well known examples include Bloch bands for electrons and phonons in periodic crystal potentials \cite{kittel} or Floquet bands for periodic-driven quantum dynamical systems \cite{grempel-84, ho-12}. If we trace the dependence of initially nondegenerate eigenvalues of a Hermitian Hamiltonian on $\bm\chi$, it is possible that, at some ${\bm \chi}$, a pair of bands  $\lambda_i({\bm \chi})$ and $\lambda_j({\bm \chi})$ goes through an exact crossing point. Such events correspond to {\it quantum phase transitions} \cite{qptbooks} that indicate, for example, a change of topological order of Floquet bands in periodic momentum space \cite{ho-12}, or electronic Bloch bands in periodic position space \cite{TI, FTI} and their phononic/magnonic counterparts \cite{zhanglifa}.  

Dissipative open systems described by non-Hermitian $\hat{H}({\bm \chi})$ also possess the  band structures  \cite{rudner-09, diehl-11}, once they have periodic dependence on position, momentum, or others. Investigation of these topological band structures is currently an active research field \cite{hu, esaki, ghosh, dissi_top}. In this Letter, we demonstrate that topological properties of the band structure of a non-Hermitian Hamiltonian can be classified with elements of the braid group. In topologically nontrivial phases such systems can undergo a spontaneous symmetry breaking and show specific detectable effects.

\begin{figure}
\scalebox{0.5}[0.5]{\includegraphics{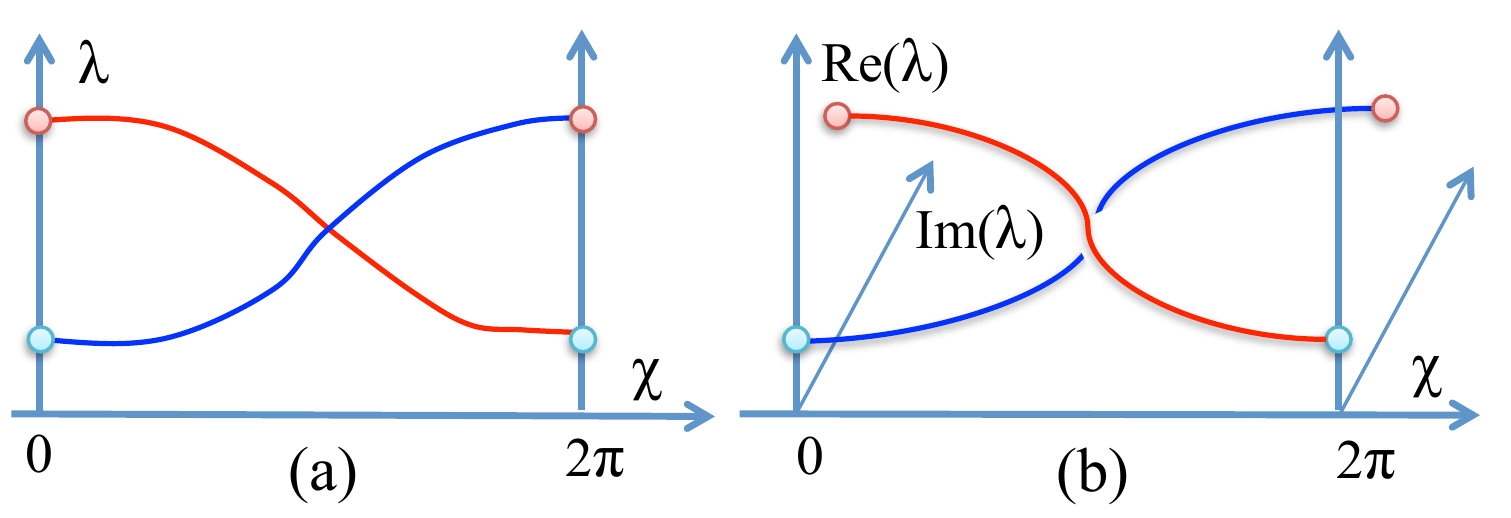}}
\hspace{-2mm}\vspace{-4mm}   
\caption{(color online). Two band structures. In order to satisfy Eqs. (\ref{period}), bands should either (a) cross or (b) create a braid pattern. The latter is possible only when eigenvalues are allowed to be complex (for non-Hermitian case).
} \label{cross}
\end{figure}

Consider first a Hermitian Hamiltonian with a periodicity $\hat H(\chi)=\hat H(\chi+2\pi)$ along one of the parameters $\chi$, and with nondegenerate eigen-spectrum $\lambda_i(\chi)$, $i=1\ldots N$. Such real eigenvalues of $\hat H(\chi)$ must have the $2\pi$-periodicity: $\lambda_i(\chi)=\lambda_i(\chi+2\pi)$. In this Letter,  we will explore consequences of the observation that in the non-Hermitian case the latter restriction generally does not hold: Eigenvalues of a non-Hermitian Hamiltonian are generally complex. Hence, the periodicity of $\hat{H}({\chi})$ only implies the periodicity of the whole nondegenerate complex eigen-spectrum but does not mean that a single eigenvalue has such a periodicity. For example, one may encounter a situation when
\begin{equation}
\lambda_i(\chi+2\pi)=\lambda_j(\chi), \quad \lambda_j(\chi+2\pi)=\lambda_i(\chi), \quad (i\ne j),
\label{period}
\end{equation}
i.e.  two complex eigenvalues, $\lambda_i$ and $\lambda_j$, twist in a braid so that the $2\pi$ periodicity for an individual eigenvalue is lost.
This situation cannot be met by nondegenerate real $\lambda_i(\chi)$ because such a case would inevitably lead to exact crossing of two eigenvalues at some $\chi$, as illustrated in Fig.~\ref{cross}(a). In contrast, Fig.~\ref{cross}(b) demonstrates that complex eigenvalues can avoid each other when  $\chi$ changes continuously while satisfying Eqs. (\ref{period}). It is straightforward to generalize this observation, namely,  the twisting pattern of $N$ band of a non-Hermitian $N\times N$ Hamiltonian as functions of the parameter $\chi$ can be generally mapped to some braid diagram of the braid group $\mathcal B_N$.  

Two band structures  are topologically equivalent if all their $N$ bands  can be continuously deformed  into each other without encountering an exact eigenvalue crossing. Different elements of the braid group cannot be continuously deformed into each other. This means that topologically equivalent bands are described by the same diagram of the braid group, i.e.   the braid group provides a natural classification of different patterns of the band structure in systems with  non-Hermitian Hamiltonians. Varying parameters of the Hamiltonian, the topology of the band structure can only change at a crossing of a degeneracy of eigenvalues. We will call such topological phase transitions the {\it braid phase transition} (BPT).

The fact that a specific band in a topologically nontrivial phase may not possess the periodicity of the Hamiltonian provides the means for the spontaneous symmetry breaking (SSB) phenomenon \cite{coleman}, \emph{i.e.}, given a symmetry of the equations of motion, a physical system can prefer certain solutions that are not invariant under this symmetry. The full symmetry is preserved only in the ensemble of all solutions. 
The SSB  is a general principle that underlies a vast number of physical phenomena, ranging from ferromagnets and superconductors  to the origin of masses in elementary particle physics \cite{nambu}. We
will show that  BPTs can lead to an unusual mechanism of the phase transition that is in some sense opposite to SSB, namely, while individual bands in a nontrivial phase do not respect the symmetry of the Hamiltonian, the observable characteristics, such as the moment generating function of currents, may keep the original symmetry. In such a case, the system may, for physical reasons, choose different solutions at different values of the periodic parameter $\chi$ , which results in discontinuity of such observables as functions of $\chi$. We will also show that, similar to Goldstone modes in SSB phases, nontrivial braid phases  correspond to the appearance of certain  modes with new properties. 

Non-Hermitian Hamiltonian generally appears in both quantum and classical open systems, once the couplings to external environments and dissipations are involved. For example, complex-valued self-energies that account for those coupling to leads inevitably appear in Green functions which when unperturbed are just based on bare Bloch band Hamiltonians. In such non-Hermitian systems, complex eigenvalues are ubiquitous, so will be the BPTs. 
As the proof of principle,  we will  explore the band structure of the operator $\hat{H}(\chi)$ that describes the transport statistics in two elementary stochastic kinetic models. $\hat{H}(\chi)$ plays the role of non-Hermitian Hamiltonian \cite{Kamenev}. The so-called counting parameter $\chi$ conjugates to the discrete transport quantities of interest, e.g., the particle number, and thus leads to the $2\pi$ periodicity of $\hat H(\chi)$, which is just similar to the periodic Brillouin zone resulting from the discrete lattice spacing in solid state physics. 

\begin{figure}[!htb]
\scalebox{0.45}[0.45]{\includegraphics{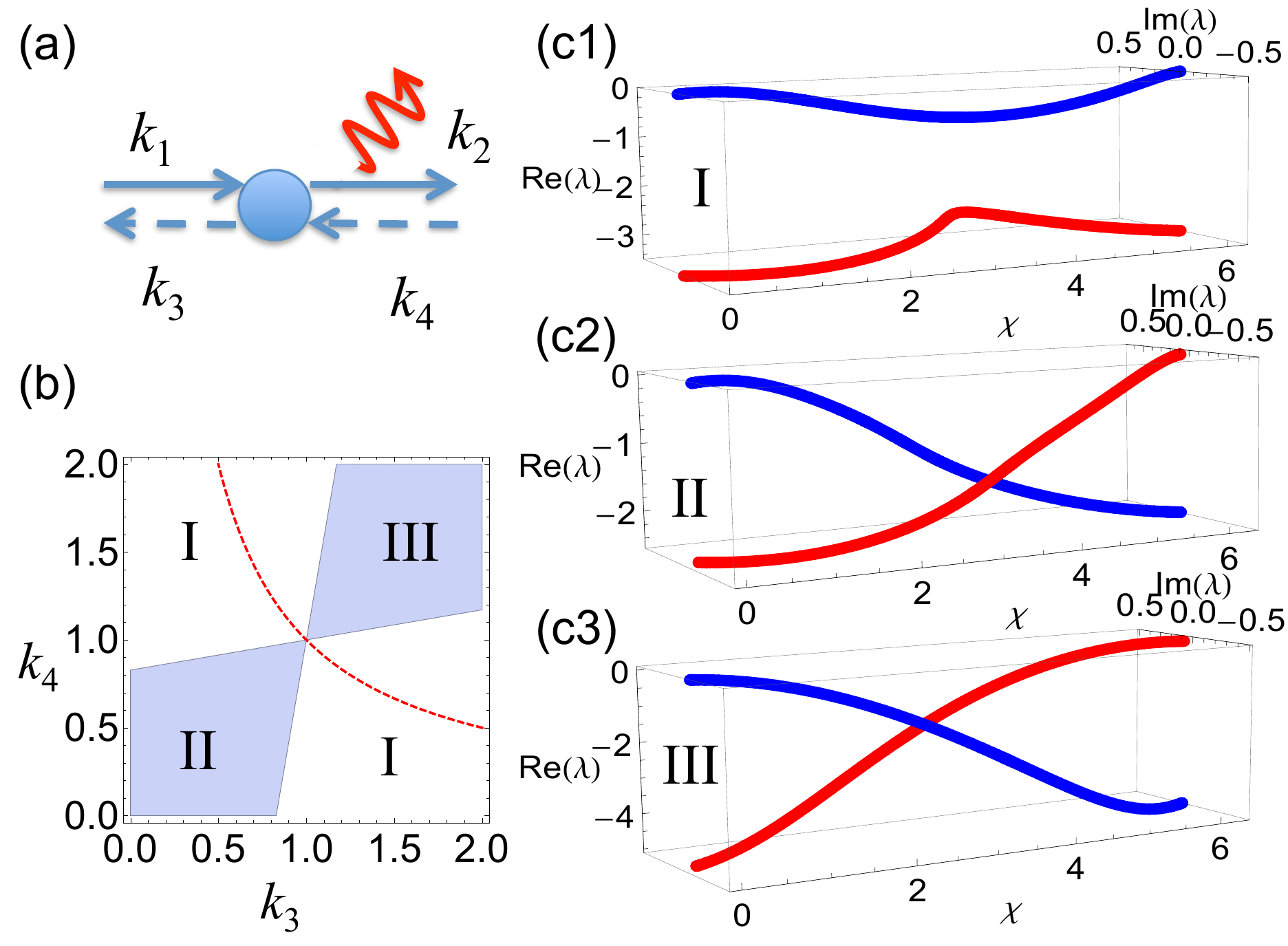}} \vspace{-8mm}
\hspace{2mm}   \caption{(color online). (a) The Michaelis-Menten-like model of a bin coupled to two particle reservoirs. 
The wavy line indicates that transitions related to the right reservoir are monitored.  (b) Braid phases of the counting statistics behavior of the kinetic model at $k_1=k_2=1$. The dashed line indicates the detailed balance condition $k_1k_2=k_3k_4$. (c) Braid diagrams of band structures for different phase regions, with (c1) phase I ($k_3=1.5, k_4=0$), 
(c2) phase II ($k_3=0.5, k_4=0$), and (c3) phase III ($k_3=1.5, k_4=1.5$).}
 \label{MM}
\end{figure}

We first consider a Markovian stochastic transport of particles between two reservoirs through an intermediate bin [see Fig.~\ref{MM}(a)]. When the bin is empty, a particle can jump from the left (right) reservoir into the bin with kinetic rate $k_1(k_4)$. When the bin has a particle inside it, the particle in the bin escapes to the left (right) reservoir with rate $k_{3}$ $(k_2)$. 
This model is known in biochemistry as the Michaeils-Menten model \cite{gopich-06, sinitsyn-07epl}, of which the rare event statistics have been measured in \cite{english-03nat}. Similar stochastic models also appear in description of quantum transports through nanoscale electric circuits \cite{pilgram-03} and phononic devices \cite{ren-10}.

Let us introduce  two sets of joint probabilities, $ p_{0}(n,t)$ and $ p_{1}(n,t)$ of that
during time interval $[0,t]$, there will be exactly $n$  particles  transferred into  the right reservoir while at time $t$ the bin is, respectively, empty or filled with a particle.
The master equations for $p_{0/1}(n,t)$ are given by
$\dot{p}_{0}(n,t) = -(k_1+k_{4})p_{0}(n,t) +k_{2}p_{1}(n-1,t)+k_{3}p_{1}(n,t)$,
$\dot{p}_{1}(n,t) = -(k_{2}+k_{3})p_{1}(n,t) +k_{1}p_{0}(n,t)+k_{4}p_{0}(n+1,t)$.
Upon a Fourier Transform, they translate into $\frac{d}{dt}(z_0, z_1)^T=\hat H(\chi)(z_0, z_1)^T$
where 
$ z_{0/1}(\chi,t)=\sum_{n=-\infty}^{\infty} p_{0/1}(n,t)e^{i\chi n}$
with the effective Hamiltonian $\hat{H}(\chi)$ given by
\begin{equation}
\hat{H}(\chi)=\left(
\begin{array}{cc}
\!\!\!\!\!\!\!\!-k_1-k_{4}& k_2e^{i\chi}+k_{3}\\
 k_{1}+k_{4}e^{-i\chi}& -k_2-k_{3}
\end{array}\right).
\label{ham2}
\end{equation}
To describe the full counting statistics of currents into the right reservoir, the moment generating function (MGF), $Z(\chi,t)=z_{0}(\chi,t)+z_{1}(\chi,t)$, can be obtained as
\begin{equation}
Z(\chi,t)
=\langle {\bm 1}|\exp\left({\hat{H}(\chi)t }\right)|{\bm p} \rangle,
\label{pdf2}
\end{equation}
where $\langle {\bm 1}|=(1,1)$, and $|{\bm p}\rangle = (p_{0},p_{1})^T$ is the vector of the initial probabilities.
The eigenvalues of $\hat{H}(\chi)$ are given by
\begin{equation}
\lambda_{\pm}(\chi)=\frac{-K\pm \sqrt{K^2+4k_1k_2e_{+}+4k_{3}k_{4}e_{-}}}{2},
\label{ev1}
\end{equation}
where $K=k_1+k_2+k_{3}+k_{4}$, $e_{\pm}\equiv e^{\pm i\chi}-1$. BPTs correspond to exact crossing of two eigenvalues, which can be found by equating the expression in the square root in (\ref{ev1}) to zero. This corresponds to the condition, $\chi=\pi$, and constraint on kinetic rates: $(k_1+k_2+k_{3}+k_{4})^2-8k_1k_2-8k_{3}k_{4}=0$. This equality describes the phase boundaries,  which divide the space of positive $k_3$ and $k_4$ into several phase regions, as shown in Fig.~\ref{MM}(b). 


In order to show that the phase transition corresponds to the topology change of the braid diagram of the band structure, we plot eigenvalues (\ref{ev1}) as a function of $\chi\in[0, 2\pi)$ in Fig.~\ref{MM}(c). Fig.~\ref{MM}(c1) corresponds to 
phase I with a trivial $2\pi$ periodicity. Fig.~\ref{MM}(c2) corresponds to phase II with eigenvalues that are not periodic but rather twisting around into each other during a continuous change of $\chi$ from $0$ to $2\pi$ and satisfying Eqs. (\ref{period}). Fig. \ref{MM}(c3) corresponds to phase III, whose braid  has a reversed chirality compared to phase II. 

The steady state current of the stochastic transport is given by ${\partial_{i\chi} \lambda_{+}(\chi)}|_{\chi=0}$, from which we see that different chiralities of the braid diagram correspond to different directions of transports. In phase II in Fig.~\ref{MM}(c2), the particle current flows from left to right while in phase III in Fig.~\ref{MM}(c3), the particle current flows from right to left as a consequence of the reversed chirality of the braid diagram. The detailed balance case, $k_1k_2=k_3k_4$, always belongs to the topologically trivial phase. It describes the equilibrium condition with zero current, which splits the phase diagram in Fig. \ref{MM}(b) into two parts with opposite current directions.

Across the phase boundaries between phase I, II, and III, the topology of the braid diagram will experience abrupt change via transitions through the degeneracy of bands.
The fact that the degeneracy point is encountered at a finite value of the counting parameter,  $\chi=\pi$, means that it influences properties of the full counting statistics rather than the lowest cumulants of a current distribution. 
In fact, at $\chi=\pi$ the matrix $\hat{H}(\pi)$ has all real entries and the corresponding MGF (We will call it the {\it parity probability})
\begin{equation}
Z(\pi,t) =   \sum_{n=-\infty}^{\infty} \sum_{j=0,1} \big[ p_j(2n,t) - p_j(2n+1,t)\big],
\label{zpi}
\end{equation}
has a simple physical meaning \cite{sinitsyn-z2}, namely, it is the difference of the probabilities to observe an even and an odd number of transitions into the right reservoir during $[0,t]$. 

\begin{figure}
\scalebox{0.35}[0.35]{\includegraphics{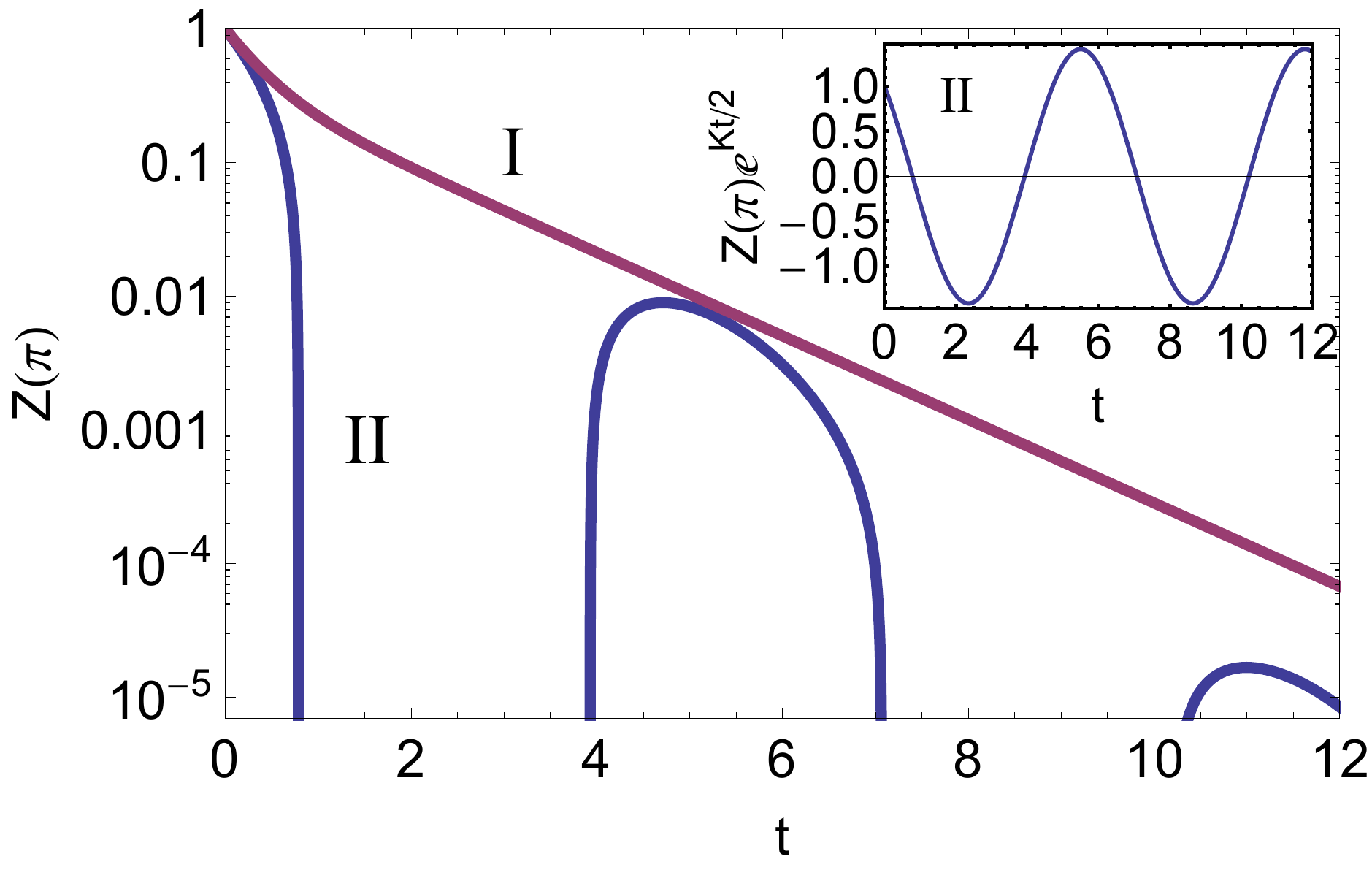}} \vspace{-5mm}
\hspace{1mm}   \caption{(color online). Manifestation of BPT by the different time evolving patterns of the parity probability $Z(\pi)$ in different phases.  Phase I ($k_{3}=1.5, k_4=0$) shows monotonic decay, while phase II ($k_{3}=k_{4}=0$) exhibits oscillatory decay. To expose oscillations of phase II, $Z(\pi)$ was multiplied by ${\rm exp}(Kt/2)$ in the inset.  
The initial condition is $|{\bm p}\rangle=(0,1)^T$.}
 \label{imre-check}
\end{figure}


The BPT directly influences the time-dependent behavior of $Z(\pi)$: Let  $\langle v_{\pm}|$ and $|u_{\pm}\rangle$ be, respectively, the left and the right eigenstates  of the matrix $\hat{H}({\pi})$, corresponding to
eigenvalues $\varepsilon_{\pm}=\lambda_{\pm}(\pi)$. Eq. (\ref{pdf2}) gives then
$Z(\pi)=C_+e^{\varepsilon_{+}t}+C_{-}e^{\varepsilon_{-}t}$, 
where $C_{\pm}=\langle {\bm 1} |u_{\pm} \rangle \langle v_{\pm}|{\bm p}\rangle$.
Since both eigenvalues $\varepsilon_{\pm}$ have negative real parts, the absolute value of $Z(\pi)$ decays exponentially with time. 
In phase I, both eigenvalues  are pure real and
$\varepsilon_+ > \varepsilon_{-}$. Hence the term $C_{-}e^{\varepsilon_{-}t}$ quickly becomes exponentially suppressed in comparison with $C_+e^{\varepsilon_{+}t}$. 
At time scales larger than $1/(\varepsilon_+-\varepsilon_{-})$ the behavior of $Z(\pi)$ is then totally described by a single exponent, as we illustrate in Fig.~\ref{imre-check}. 

Consider next the case of parameters in phase II. Two eigenvalues, $\varepsilon_{\pm}$, are now complex conjugate to each other. They have the same real part, which means that it is impossible to disregard one solution in favor of another one. Moreover, it is easy to see that $C_{-}=C_{+}^*$ and the time-evolution of the parity probability becomes oscillating 
\begin{equation}
Z(\pi)=Ae^{-Kt/2} {\rm cos}({\rm Im}(\varepsilon_+)t +\phi_0),
\label{stat2}
\end{equation}
with some constants $A$ and $\phi_0$ that depend on initial conditions. 
In Fig.~\ref{imre-check}, we show numerical results of the evolution of the parity probability in phase II, which confirm that the BPT to the topologically nontrivial phase corresponds to oscillations of $Z(\pi)$. This also implies that the conventional assumption of sufficiency of only one eigenvalue to estimate the full counting statistics at long times cannot be used in the nontrivial braid phase II. Phase III shares similar behaviors. 

\begin{figure}
\scalebox{0.4}[0.4]{\includegraphics{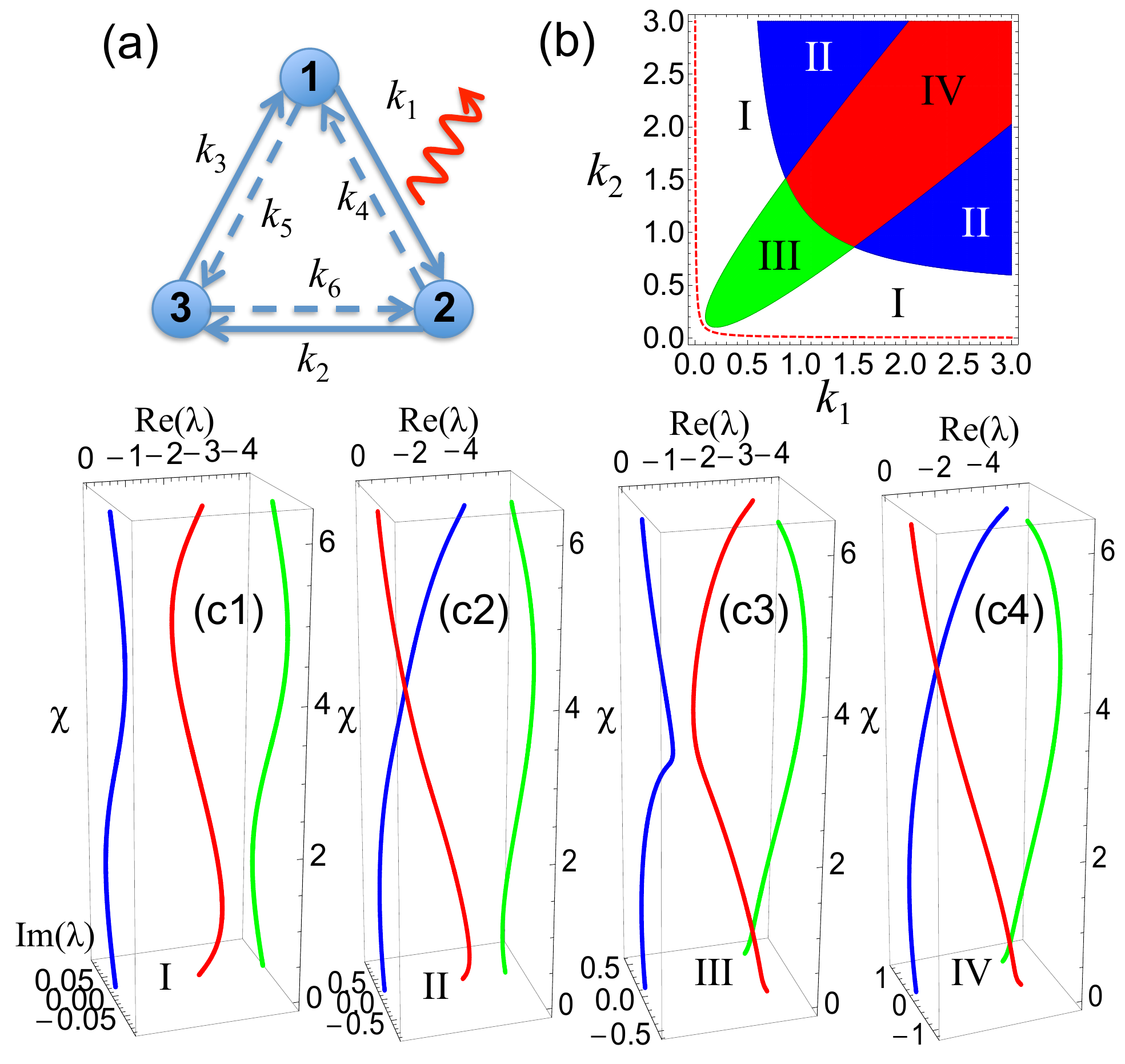}} \vspace{-5mm}
\hspace{2mm}  \caption{(color online). (a) Scheme of three-state cyclic kinetics. 
(b) The phase diagram of the transport statistics of the three-state model. The dashed line indicates the detailed balance condition $k_1k_2k_3=k_4k_5k_6$ at equilibrium \cite{note_neq}. (c) Braid diagrams of band structures for different phases: (c1) I with $k_1= 2, k_2=0.1$; (c2) II $(k_1= 1, k_2 = 3)$; (c3) III $(k_1= 1, k_2= 1)$. (c4) IV $(k_1= 2, k_2= 2)$. }
 \label{3S} 
\end{figure}

As a more complex example, we consider next a Markovian three-state kinetic model shown in Fig. \ref{3S}(a). It is widely used as a model for cyclic enzyme reactions \cite{qian-review},  molecular motors \cite{astumian},  and charge currents through quantum dots \cite{Utsumi}. With transitions between states $1$ and $2$ being monitored, the evolution of the counting statistics is described by
$\frac{d}{dt} (z_1, z_2, z_3)^T=\hat H(\chi)(z_1, z_2, z_3)^T$,
where the effective Hamiltonian $\hat{H}(\chi)$ is 
\begin{equation}
\hat{H}(\chi)=\left(
\begin{array}{ccc}
-k_1-k_5  & k_4 e^{-i\chi} & k_3 \\
 k_1e^{i\chi} & -k_2-k_4 & k_6  \\
 k_5 & k_2 & -k_3-k_6 
\end{array}
\right),
\label{ham3}
\end{equation}
 $p_j(n,t)$ is the joint probability 
for a system to be in state $j$ while there are already $n$ transitions from state $1$ to $2$ observed by  time $t$.

The system now has three bands $\lambda_{j=1,2,3}(\chi)$ and a richer braid phase diagram.
For the sake of simplicity,  we fix $k_3=2.1, k_4=2, k_5=0.1, k_6=0.1$ and only plot the phase diagram in the space of $(k_1, k_2)$ in Fig. \ref{3S}(b). There are four braid phase regions \cite{footnote1}. Phase I is the trivial braid with each band keeping the $2\pi$ period [see Fig. \ref{3S}(c1)]. Phase II can be developed from phase I by crossing $\lambda_1(\chi)$ and $\lambda_2(\chi)$ at $\chi=\pi$. The two new bands twist and transfer into each other after a continuous change of $\chi$ by $2\pi$, with the third band keeping its periodicity intact [see Fig. \ref{3S}(c2)]. The nontrivial topology of bands is manifested by the decaying oscillation of the parity probability $Z(\pi)=\sum^{3}_{j=1}z_j(\pi)$.

In phase III, the situation is different. The first band remains intact while the second and third bands are twisting so that $\lambda_2(\chi+2\pi)=\lambda_3(\chi)$ and $\lambda_3(\chi+2\pi)=\lambda_2(\chi)$ [see Fig. \ref{3S}(c3)]. This braid phase can be reached from phase I via the degeneracy of $\lambda_2(\chi)$ and $\lambda_3(\chi)$ at $\chi=0$, which then splits into a conjugate eigenvalue pair. Since at $\chi=0$, $z_j(\chi=0)$ denotes the conventional probability of the system to be in state $j$, this new topological phase corresponds to the emergence of decaying oscillations of conventional state probabilities to their steady states.

Phase IV is a combination of phase II and III, in which three bands are twisted around each other. Each band becomes periodic in $\chi$ with a period equal to $6\pi$ [see Fig. \ref{3S}(c4)]. The observable effect of this braid phase is also a combination of decaying oscillations of both the conventional state probabilities and the parity probability $Z(\pi)$.

Therefore, oscillating modes for the parity and state probability can be used to identify the nontrivial braid topology. Opposite flux directions can be further used to distinguish different braid phases with different chirality. 
Combinations of the different transport directions and the oscillating behaviors of different probabilities are able to reveal the different braid phases. 
When beyond the three-band model, the strategies are similar and the phenomena are combinations of opposite flux directions, oscillating modes of the conventional state probabilities and the parity probabilities. They are fortunately detectable by today's measuring techniques \cite{english-03nat, Utsumi, Nakamura, measuring}.

 

Finally, we would like to discuss the non-analytic behavior of the MGF near $\chi=\pi$, which has also encountered in other contexts previously \cite{sinitsyn-z2,pistolesi,abanov-z2,abanov-11,mirlin-08prb}. In particular, in quantum 1D electron systems it was observed that the analytical continuations of generating functions at $\chi=0$ and $2\pi$ do not coincide near $\chi=\pi$  \cite{mirlin-08prb}. It was proved recently \cite{abanov-11} that near $\chi =\pi$ the generating function should be given by a sum of solutions. 
In our model, this fact has a simple explanation. The parity probability in phase II is, indeed,  given by a sum of two equal in absolute value contributions because two eigenvalues of $\hat{H}(\pi)$ have equal real part in this phase.  In phase II, an individual eigenvalue near $\chi = 0$ cannot be uniquely continued to the point $\chi = \pi$ because the square root in  (\ref{ev1})  is doubly valued. At the steady state and at $\chi\neq\pi$, the system chooses the state with the largest real part of the eigenvalue, which is a part of one band at $\chi\in[0,\pi)$ and a part of another band at $\chi\in(\pi,2\pi]$, and thus leads to the discontinuity at $\chi=\pi$. By summing the two band branches, the MGF restores the broken symmetry.

In equilibrium statistical physics, a phase transition can be identified by the non-analytic singularity of free energy, {\it e.g.}, Lee-Yang zeros \cite{LeeYang1}. In nonequilibrium transports, the MGF $Z(\chi,t)$ of dynamical observables is the counterpart of the partition function and $\ln Z(\chi)$ is the counterpart of the free energy. Thus, the zeros in $Z(\chi)$ and their time-dependence play a similar role to the Lee-Yang zeros and can be used to explore the dynamical phase transitions \cite{rare-review, garrahan, qian, LeeYang2, LeeYang3}. Indeed, as shown by Eq. (\ref{stat2}), in the nontrivial braid phase region the parity probability $Z(\pi)$ oscillates and its zeros as functions of time are one-to-one mapped to the Lee-Yang zeros of the stochastic transport system.


In summary, we have shown that, when parameters change, the band structures of non-Hermitian Hamiltonians may encounter topological phase transitions that can be classified by the elements of the braid group. In application to the full counting statistics of currents, we found that topologically nontrivial phases correspond to oscillating behavior of parity and state probabilities. They also correspond to discontinuities of the MGF of currents induced via selection of different band branches in the limit of long time evolution. 

For bands of evolution operator of a periodically driven quantum or classical system, the eigenvalues are generally complex and can possess cyclic permutations as a function of Bloch-Floquet vectors. The BPTs can be implicated in such time-dependent driven systems and play a crucial role in understanding their dynamical behaviors and related dynamical phase transitions.
Open problems include generalizations of BPTs to Hamiltonians that depend on a multi-component Bloch vector and other manifestations of BPTs in quantum mechanical full counting statistics.

\begin{acknowledgments}
{This work was supported by the National Nuclear Security Administration of the U.S. DOE at LANL under Contract No. DE-AC52-06NA25396, and the LDRD Program at LANL.}
\end{acknowledgments}


\begin{thebibliography}{300}
\bibitem{note_neq} It is worth remarking that out-of-equilibrium is the necessary condition for the BPT because when detailed balance is respected, the eigen-spectra of the non-Hermitian Hamiltonian with counting parameters are always real. To have nontrivial braids, the detailed balance condition at equilibrium must be broken.
\bibitem{footnote1} In fact, if we explore all possible parameters, there are three more phase regions with reversed chirality as well as reversed transport direction compared to the braid phase II, III and IV, respectively.
\end{thebibliography}

\end{document}